\documentclass{iopjournal}

\usepackage{graphicx, dblfloatfix}
\usepackage{hyperref}
\usepackage{cite}
\usepackage{color, xcolor}
\usepackage{rotating}
\usepackage{tabularx}

\usepackage{booktabs}
\usepackage{multirow}
\usepackage{makecell}   
\usepackage{amssymb}
\usepackage{braket}
\usepackage{verbatim}
\usepackage{verbatim}
\usepackage{float}

\expandafter\let\csname equation*\endcsname\relax
\expandafter\let\csname endequation*\endcsname\relax
\usepackage{amsmath}

\newcommand{\trm}{\textrm}

\usepackage{lineno}
\usepackage[T1]{fontenc}

\begin{document}

\articletype{Paper} 

\title{Real-time feedback control of ELM frequency using divertor gas puffing and its effects on tungsten-induced radiation and plasma performance in KSTAR}

\author{Minseok Kim$^{1}$\orcid{0009-0002-1664-6618}, Young-Ho Lee$^2$\orcid{0000-0002-5285-4599}, SangKyeun Kim$^3$\orcid{0000-0002-0701-8962}, Minwoo Kim$^2$\orcid{0000-0002-8627-4584}, Sang-hee Hahn$^2$\orcid{0000-0001-8115-9248}, Hiro J. Farre-Kaga$^{1,3}$\orcid{0009-0004-3640-4557}, Ricardo Shousha$^3$\orcid{0000-0003-1498-8980}, Juhyeok Jang$^2$\orcid{0000-0002-9991-6072}, SooHyun Son$^2$\orcid{0000-0001-8271-1495}, Yoon Seong Han$^4$\orcid{0009-0004-3018-3837}, Junghoo Hwang$^4$\orcid{0000-0001-5891-6757}, Boseong Kim$^2$\orcid{0009-0002-0109-6502}, and Egemen Kolemen$^{1,3,*}$\orcid{0000-0003-4212-3247}}

\affil{$^1$Department of Mechanical and Aerospace Engineering, Princeton University, Princeton, NJ 08540, USA}

\affil{$^2$Korea Institute of Fusion Energy (KFE), Daejeon 34133, Republic of Korea}

\affil{$^3$Princeton Plasma Physics Laboratory, Princeton, NJ, 08540, USA}

\affil{$^4$Department of Nuclear and Quantum Engineering, Korea Advanced Institute of Science and Technology (KAIST), Daejeon 34141, Republic of Korea}

\affil{$^*$Author to whom any correspondence should be addressed.}

\email{mseokim@princeton.edu and ekolemen@princeton.edu}


\begin{abstract}
The edge-localized mode (ELM) frequency ($f_{\mathrm{ELM}}$) was successfully controlled in real time on KSTAR using a proportional-integral (PI) feedback controller, employing a $\mathrm{D}_2$ divertor gas puff as the actuator under tungsten lower-divertor conditions. The controller accurately tracked a two-step target---a 30 Hz increase in $f_{\mathrm{ELM}}$ for 4 s, followed by a 30 Hz decrease for 3 s---yielding mean and median absolute percentage errors of approximately 13$\%$ and 12$\%$, respectively. Compared to a reference discharge, the actively controlled shot did not exhibit a significant drop in volume-integrated core radiation, confirming that excessive gas use merely degrades overall plasma performance. However, when contrasted with the exponential increase in core radiation observed in the absence of divertor gas puffing, these results underscore the critical need for real-time optimization. Specifically, divertor gas commands must be actively managed to maintain an $f_{\mathrm{ELM}}$ sufficient for flushing tungsten from the core while maximizing global plasma performance.
\end{abstract}

\section{Introduction}
Tungsten (W) is considered a highly promising candidate for tokamak plasma-facing components due to its higher melting point and lower tritium retention compared to carbon or beryllium \cite{paper:philipps2011tungsten, paper:ueda2014research}. Its operational feasibility has been extensively investigated across various machines, including JET\cite{paper:matthews2011jet, paper:romanelli2011overview, paper:romanelli2015overview, paper:litaudon2017overview}, WEST\cite{paper:bucalossi2014west, paper:bucalossi2022operating, paper:bucalossi2026overview, paper:hakola2026overview}, ASDEX-U\cite{paper:neu2007operational, paper:neu2013overview, paper:stroth2013overview, paper:kallenbach2017overview}, EAST\cite{paper:zhou2015upgrade, paper:cao2015east, paper:yao2016design, paper:xu2021physics}, DIII-D\cite{paper:abrams2021design, paper:abrams2023recent, paper:turco2024first, paper:cappelli2026analysis}, and KSTAR \cite{paper:kim2025thermal, paper:ahn2025tungsten, paper:kim2025vacuum, paper:kwon2025engineering, paper:kwak2026study, paper:nam2026overview}. Despite its thermomechanical advantages, tungsten is a high-Z material; its accumulation in the plasma core drives strong radiation, which degrades overall performance and can trigger plasma disruptions. Consequently, minimizing the core tungsten concentration is essential for stable, high-performance operation. Edge-localized modes (ELMs), which periodically expel particles and energy from the plasma edge \cite{paper:zohm1996edge, paper:connor1998edge, paper:leonard2014edge}, are known to effectively flush impurities from the plasma \cite{paper:burrell1989confinement, paper:pitcher1997experimental, paper:dux2009plasma, paper:dux2011main, paper:maggi2015pedestal}. Maintaining the ELM frequency ($f_{\mathrm{ELM}}$) above a certain threshold is therefore necessary to prevent tungsten accumulation. Because gas puffing is an effective method for controlling $f_{\mathrm{ELM}}$ \cite{paper:maggi2015pedestal, paper:lennholm2015elm, paper:lennholm2016real}, real-time divertor gas management is critical. This necessity was clearly demonstrated on KSTAR (using a bottom piezoelectric valve with $D_2$ gas, PVB): terminating the divertor gas puff resulted in sporadic giant ELMs and an exponential increase in core radiation, ultimately leading to a disruption, as shown in Figure \ref{fig:ELMDetection}.

To further study the effect of $f_{\mathrm{ELM}}$ on tungsten-induced radiation and plasma performance, we developed a proportional-integral (PI) feedback loop to dynamically control $f_{\mathrm{ELM}}$ in real time using a $\mathrm{D}_2$ divertor gas puff as the actuator on KSTAR. The ELM frequency response to the divertor gas puff, shown in Figure \ref{fig:ELMDetection}, was used to design the PI controller. The controller successfully tracked a target of $f_{\mathrm{ELM}} = 140$ Hz. This was compared against a reference discharge that utilized a 1.5 V feedforward $D_2$ divertor gas puff (DH port) yielding $f_{\mathrm{ELM}} \sim 100$ Hz. The two discharges were compared across several parameters: line-averaged density ($\bar{n}_\mathrm{e}$) measured by the two-color interferometer (TCI) \cite{paper:lee2016design, paper:juhn2021multi, paper:juhn2026fpga}, volume-integrated core ($P_{\mathrm{core}}$, $\psi_{\mathrm{N}} \leq 0.8$) and edge ($P_{\mathrm{edge}}$, $\psi_{\mathrm{N}} > 0.8$) radiation measured by bolometry \cite{paper:seo2010first, paper:jang2018tomographic, paper:jang2018reconstruction, paper:oh2018forward, paper:oh2024data, paper:han2026high}, ion temperature ($T_{\mathrm{i}}$) from the charge exchange system (CES) \cite{paper:ko2010charge, paper:ko2010kstar, paper:lee2010development, paper:oh2011optical, paper:lee2011development, paper:Lee2020TwoGaussian, paper:jang2021development, paper:lee2022measurement, paper:lee2026neural}, and normalized toroidal beta ($\beta_{\mathrm{N}}$) reconstructed via EFIT \cite{paper:EFIT, paper:jeon2010equilibrium, paper:lee2011equilibrium}. This comparison reveals that using excessive divertor gas to increase $f_{\mathrm{ELM}}$ elevated both $\bar{n}_\mathrm{e}$ and $P_{\mathrm{edge}}$, but degraded $T_{\mathrm{i}}$ and $\beta_{\mathrm{N}}$, while only marginally reducing $P_{\mathrm{core}}$ within the measurement uncertainty limit. These results emphasize the critical need for real-time divertor gas optimization to balance core W flushing with optimal plasma performance.

The remainder of this paper is organized into two main parts: controller design and experimental results. Section \ref{sec:controller} details the real-time calculation of $f_{\mathrm{ELM}}$ and the design of the PI controller. To facilitate feedback design, the plasma response to decreased divertor gas puffing is approximated using a first-order-plus-dead-time (FOPDT) model. The control strategy is illustrated using Bode and Nyquist plots, while the closed-loop poles, zeros, and bandwidths are detailed in \ref{appendix:control}. Section \ref{sec:control_results} presents the experimental control results (shot 43187) compared against a reference discharge (shot 43186). During the experiment, $f_{\mathrm{ELM}}$ was successfully regulated at 140 Hz from $t = 6$ to 10 s, followed by a reduction to 110 Hz until $t = 13$ s, by adaptively applying the divertor gas puff. The subsequent effects on $\bar{n}_\mathrm{e}$, $P_{\mathrm{edge}}$, $T_{\mathrm{i}}$, and $\beta_{\mathrm{N}}$ are analyzed. Additionally, two-dimensional (2D) profiles of $P_{\mathrm{rad}}$ and $T_{\mathrm{i}}$ are compared across three distinct phases: $t = 6.00$ s (prior to the $f_{\mathrm{ELM}}$ increase), $t = 8.52$ s (during the elevated $f_{\mathrm{ELM}}$ phase), and $t = 12.0$ s (after the $f_{\mathrm{ELM}}$ reduction). Finally, Section \ref{sec:conclusion} summarizes the primary conclusions.

\section{Real-time \texorpdfstring{f$_{\mathrm{ELM}}$}{f\_ELM} detection and a feedback controller design}
\label{sec:controller}

\begin{figure}[tbp]
\begin{center}
\includegraphics[width= \linewidth]{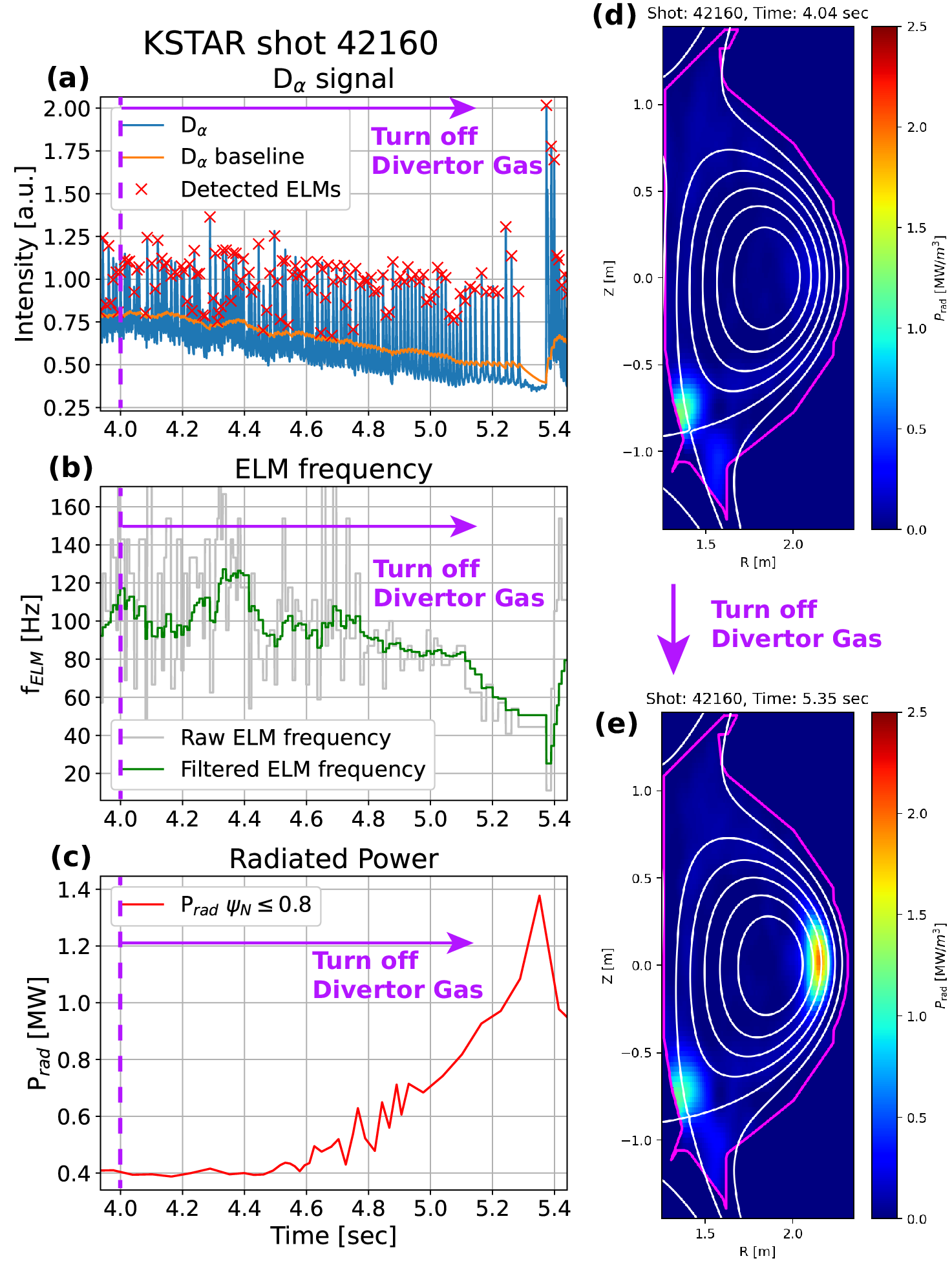}
\caption{
Evolution of $f_{\mathrm{ELM}}$ and $P_{\mathrm{rad}}$ for KSTAR shot 42160, in which the divertor gas puff was intentionally terminated at $t = 4$ s. \textbf{(a)} Time traces of the $\mathrm{D}_\alpha$ signal, its extracted baseline, and the detected ELM peaks. The raw signal is processed using the low-pass filter (LPF) defined in Eq. (\ref{eq:LPF}) with $\tau = 455\ \mu\mathrm{s}$ (cut-off frequency $\sim 300$ Hz), while the baseline is extracted using $\tau = 0.05$ s. ELM spikes are identified from the baseline-subtracted signal using empirical thresholds for gradient ($> 200$), height ($> 0.2$), and the presence of a subsequent negative gradient. \textbf{(b)} Calculated ELM frequency, $f_{\mathrm{ELM}} = 1 / \Delta t_{\mathrm{ELM}}$, where $\Delta t_{\mathrm{ELM}}$ is the time interval between consecutive ELM peaks. The plotted $f_{\mathrm{ELM}}$ trace is smoothed using an LPF with $\tau = 0.05$ s. \textbf{(c)} Volume-integrated core radiated power ($P_{\mathrm{core}}$) within the $\psi_{\mathrm{N}} \leq 0.8$ surface. \textbf{(d)} Two-dimensional (2D) $P_{\mathrm{rad}}$ profile at $t = 4.04$ s, immediately following the divertor gas termination. \textbf{(e)} 2D $P_{\mathrm{rad}}$ profile at $t = 5.35$ s, exhibiting strong core radiation driven by W accumulation approximately 1.35 s after the gas puff was turned off.
}
\label{fig:ELMDetection}
\end{center}
\end{figure}

High-frequency (20 kHz) visible spectroscopy, measuring deuterium Balmer-alpha ($D_{\alpha}$) intensity at the plasma edge \cite{seminar:Jang2025visible}, captures the occurrence of ELMs as distinct spikes in the signal. It allows the ELM frequency ($f_{\mathrm{ELM}}$) to be calculated as the inverse of the time interval between consecutive spikes. An ELM peak is identified when both the signal amplitude and its gradient exceed predefined thresholds. To accurately evaluate these criteria, the signal baseline is first subtracted, and a first-order low-pass filter (LPF) is applied to suppress noise. The discrete-time LPF is given by

\begin{align}
y_{\mathrm{LPF}}(t) = y_{\mathrm{LPF}}(t-1) + \frac{\Delta t}{\Delta t + \tau} \left[ y(t) - y_{\mathrm{LPF}}(t-1) \right], \label{eq:LPF}
\end{align}

\noindent where $\Delta t = 500\ \mu\mathrm{s}$ is the cycle time of the real-time CPU, running the $f_{\mathrm{ELM}}$ detection and control algorithm, in the KSTAR plasma control system (PCS). An LPF with the time constant of $\tau_{\mathrm{base}} = 0.05$ s is used to extract the moving baseline. The baseline is subtracted from the signal, then filtered using $\tau_{D_{\alpha}} = 455\ \mu\mathrm{s}$, which corresponds to a cut-off frequency of $\sim 300$ Hz. The empirical gradient and height thresholds for ELM detection are set to 200 [a.u.] and 0.2 [a.u.], respectively. Upon detecting an ELM, $f_{\mathrm{ELM}}$ is updated as $1 / \Delta t_{\mathrm{ELM}}$, where $\Delta t_{\mathrm{ELM}}$ is the time interval between the current and previous peaks. To mitigate the impact of false detections, the calculated $f_{\mathrm{ELM}}$ is subsequently filtered using an LPF with $\tau_{f_{\mathrm{ELM}}} = 0.05$ s. This time constant is small enough to capture the dynamic changes in $f_{\mathrm{ELM}}$ driven by the divertor gas puffing. Figure \ref{fig:ELMDetection} (a) illustrates the $D_{\alpha}$ baseline alongside the detected ELM spikes, while the resulting $f_{\mathrm{ELM}}$ trace is shown in Figure \ref{fig:ELMDetection} (b).

To evaluate the $f_{\mathrm{ELM}}$ response to the divertor gas puff, KSTAR shot 42160 (Figure \ref{fig:ELMDetection}) was selected. This is a standard lower single-null H-mode discharge designed to avoid core W accumulation via divertor gas puffing. The plasma parameters are: $I_{\mathrm{p}} = 0.60$ MA, $B_{\mathrm{T}} = 1.9$ T, $q_{95} \sim 4.6$, $\beta_{\mathrm{N}} \sim 2.2$, $l_{\mathrm{i}} \sim 0.84$, $\kappa \sim 1.8$, $\delta_{\mathrm{upper}} \sim 0.32$, and $\delta_{\mathrm{lower}} \sim 0.84$. Heating and fueling include $\sim 5.0$ MW of neutral beam (NB) injection, $\sim 0.50$ MW of electron cyclotron (EC) heating, and a main gas puff of $\sim 1$ V up to $t = 2$ s (with a 0.6 V prefill). While the standard scenario applies a 3 V divertor gas puff from $t = 1$ to 4 s followed by a 1.5 V puff, we intentionally terminated the divertor gas at $t = 4$ s in shot 42160 to observe the $f_{\mathrm{ELM}}$ response. As shown in Figures \ref{fig:ELMDetection} (a) and (b), shutting off the divertor gas reduces both the overall $\mathrm{D}_{\alpha}$ baseline signal and $f_{\mathrm{ELM}}$. Consequently, this triggers an exponential increase in $P_{\mathrm{core}}$, as depicted in Figure \ref{fig:ELMDetection} (c). This dynamic response is characterized using a first-order-plus-dead-time (FOPDT) model; the derivation of the model coefficients and the controller design strategies are detailed in \ref{appendix:control}.

\section{\texorpdfstring{f$_{\mathrm{ELM}}$}{f\_ELM} control results}
\label{sec:control_results}

\begin{figure}[tbp]
\begin{center}
\includegraphics[width = \linewidth]{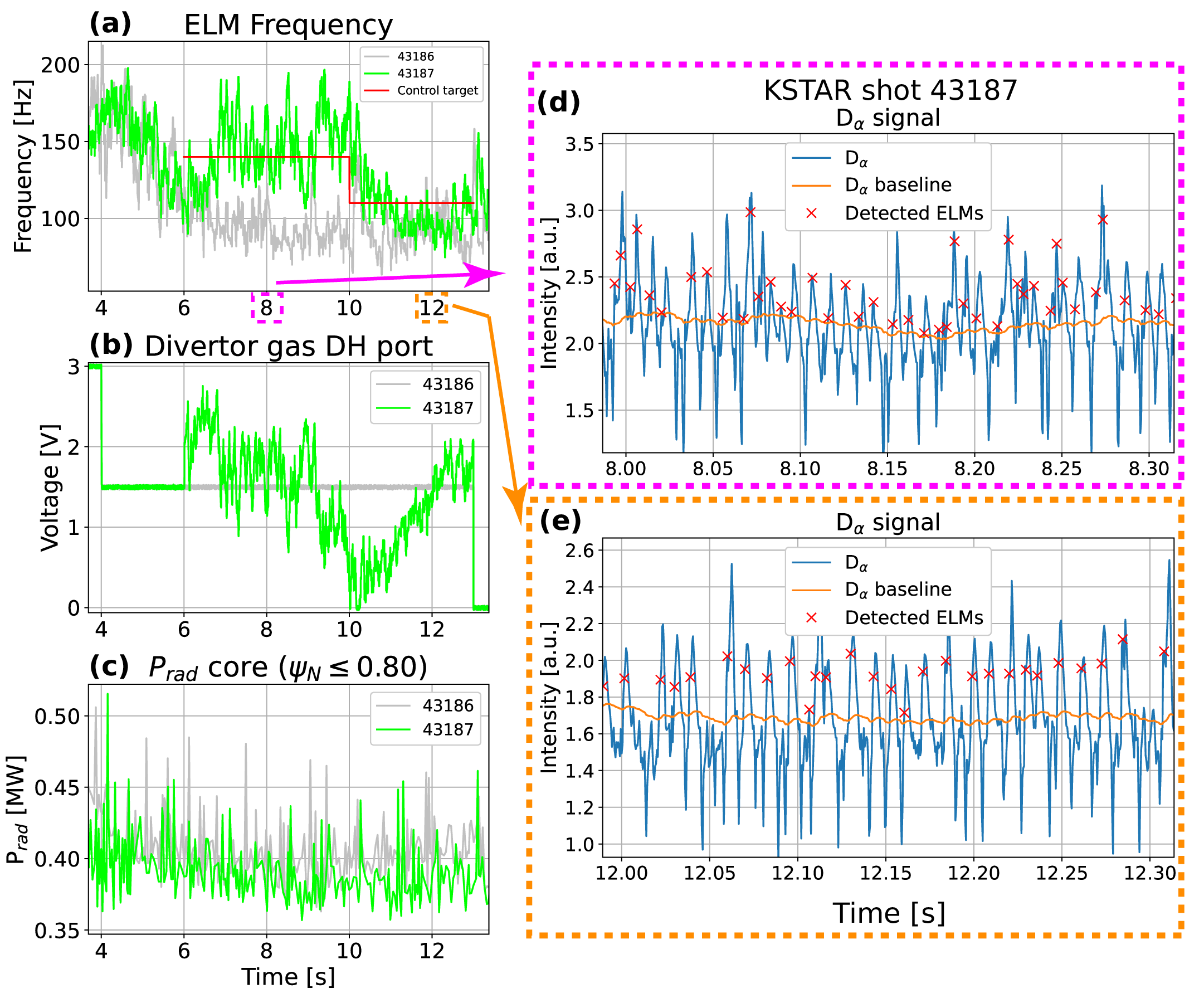}
\caption{
Comparison between the reference discharge (KSTAR shot 43186, gray) and the controlled discharge (KSTAR shot 43187, lime). \textbf{(a)} Evolution of $f_{\mathrm{ELM}}$, where the red line indicates the real-time control target ($140$ Hz from $t = 6$ to $10$ s, followed by $110$ Hz until $t = 13$ s). \textbf{(b)} Voltage command applied to the divertor gas puff at the DH port. \textbf{(c)} Volume-integrated core radiated power ($P_{\mathrm{core}}$) within the $\psi_{\mathrm{N}} \leq 0.8$ surface. Detected ELM peaks and extracted $\mathrm{D}_{\alpha}$ baseline signals are shown in detail at \textbf{(d)} $t = 8.00$ s and \textbf{(e)} $t = 12.00$ s.
}
\label{fig:control}
\end{center}
\end{figure}

Shot 43187 was dedicated to the dynamic control of $f_{\mathrm{ELM}}$ and was compared against a reference discharge, shot 43186. The reference discharge utilized the same $I_{\mathrm{p}}$, $B_{\mathrm{T}}$, and NB heating power as shot 42160 (used for the controller design in Figure \ref{fig:ELMDetection}), but a 1.5 V divertor gas puff was applied starting at $t = 4$ s. Due to fallen wall debris in the divertor at the time, the outer strike point was shifted inward relative to shot 42160 to avoid hitting the material, resulting in a higher $\delta_{\mathrm{lower}}$. The EC heating and prefill gas configurations were also fine-tuned for this experiment. Furthermore, a boron dropper was operated at 4 V from $t = 14$ to 16 s to maintain wall conditions, compensating for the temporary unavailability of the in-vessel cryopump (IVCP) during the week of the experiment. In shot 43187, active control was engaged at $t = 6$ s, with the $f_{\mathrm{ELM}}$ target set to 140 Hz from $t = 6$ to 10 s, followed by a step down to 110 Hz until $t = 13$ s. As shown in Figure \ref{fig:control} (b), the PI controller adaptively modified the divertor gas puff commands to accurately track the $f_{\mathrm{ELM}}$ target (Figure \ref{fig:control} (a)). The controller successfully followed the target with mean and median absolute percentage errors of 13.4 \% and 11.6 \%, respectively, over the active control window ($t = 6$ to 13 s). A comparison of $P_{\mathrm{core}}$ with and without $f_{\mathrm{ELM}}$ control (Figure \ref{fig:control} (c)) reveals an approximate 5 \% reduction in core radiation at $t = 10$ s; however, this small drop falls within the margin of diagnostic uncertainty.

\begin{figure}[tbp]
\begin{center}
\includegraphics[width = 0.45 \linewidth]{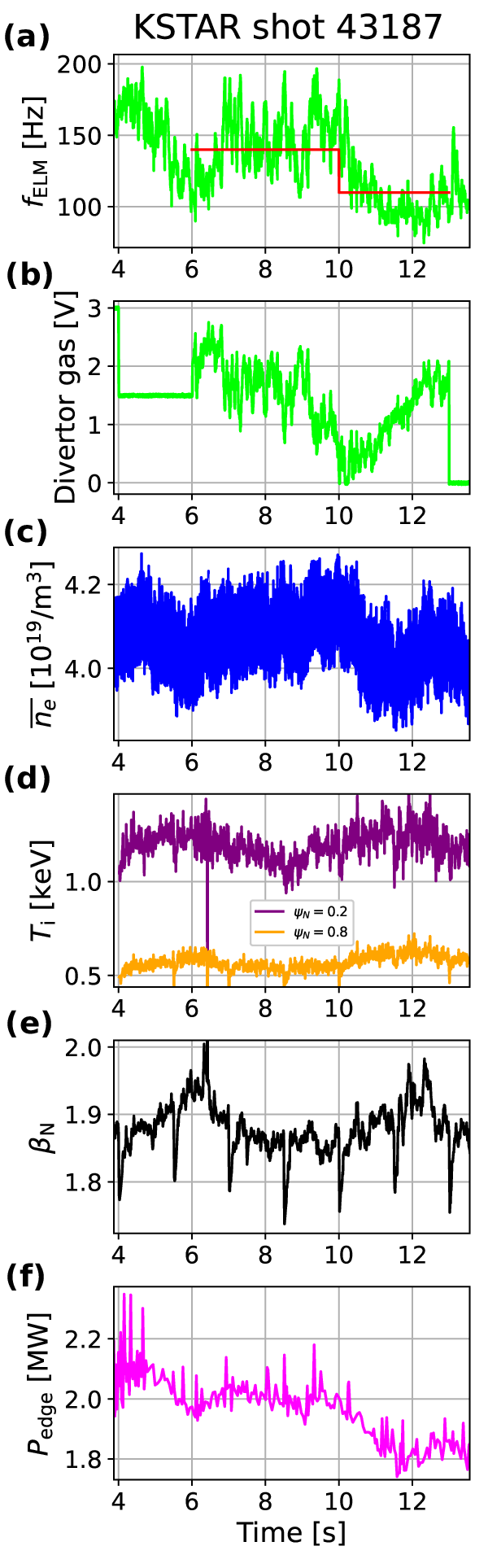}
\caption{
Time traces of plasma parameters and actuator commands for the actively controlled KSTAR shot 43187. \textbf{(a)} Measured $f_{\mathrm{ELM}}$ against the real-time control target. \textbf{(b)} Divertor gas puff voltage command applied at the DH port. \textbf{(c)} Line-averaged electron density ($\bar{n}_\mathrm{e}$) measured by TCI channel 1, which represents the longest line of sight among the five channels. \textbf{(d)} Ion temperature ($T_{\mathrm{i}}$) at $\psi_{\mathrm{N}} = 0.2$ and 0.8, measured by the charge exchange system (CES). \textbf{(e)} Normalized toroidal beta ($\beta_{\mathrm{N}}$). \textbf{(f)} Volume-integrated edge radiated power outside the $\psi_{\mathrm{N}} = 0.8$ surface. The $T_{\mathrm{i}}$ profiles shown in \textbf{(d)} were fitted using Gaussian process regression \cite{paper:ms_GPR, book:Rasmussen, book:murphy, paper:kwak, paper:chilenski2015improved, paper:ho2019application}.
}
\label{fig:kinetics}
\end{center}
\end{figure}

\begin{figure}[tbp]
\begin{center}
\includegraphics[width = \linewidth]{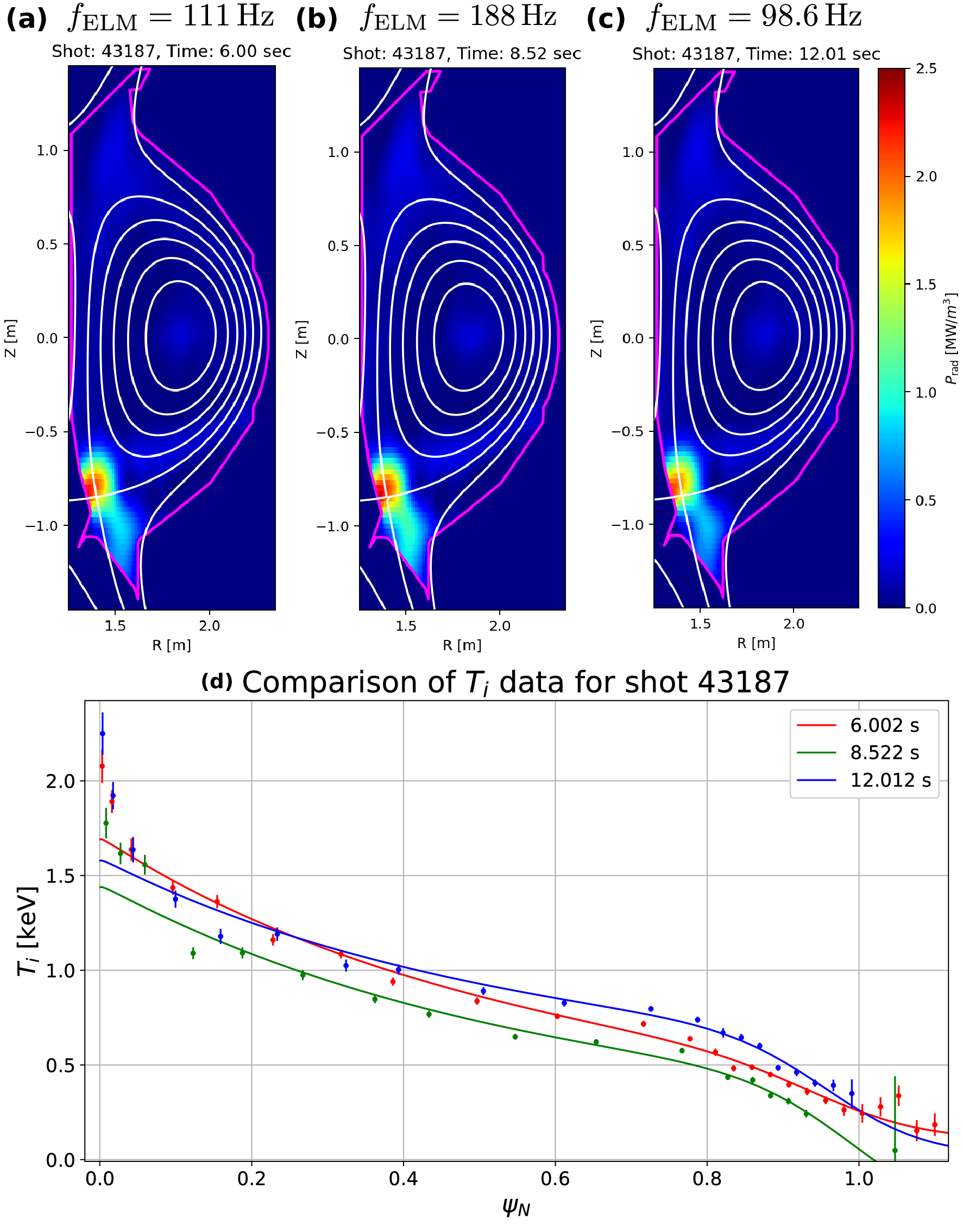}
\caption{
Comparison of two-dimensional (2D) radiated power density [MW/m$^3$] and $T_{\mathrm{i}}$ profiles [keV] across three time slices during the control experiment. 2D radiation profiles are shown \textbf{(a)} at $t = 6.00$ s (control onset), \textbf{(b)} during the high-$f_{\mathrm{ELM}}$ control phase, and \textbf{(c)} following the $f_{\mathrm{ELM}}$ ramp-down. \textbf{(d)} Evolution of the 1D $T_{\mathrm{i}}$ profile across these three time slices. The increased gas puff resulted in a higher $f_{\mathrm{ELM}}$, enhanced radiation in the divertor region, and an overall reduction in $T_{\mathrm{i}}$.
}
\label{fig:rad_Ti}
\end{center}
\end{figure}

As shown in Figure \ref{fig:kinetics}, increasing the divertor gas puff voltage to control $f_{\mathrm{ELM}}$ elevated $\bar{n}_\mathrm{e}$ and reduced $T_{\mathrm{i}}$, consequently causing a drop in $\beta_{\mathrm{N}}$. Furthermore, this gas injection enhanced $P_{\mathrm{rad}}$ near the divertor (Figures \ref{fig:kinetics} (f) and \ref{fig:rad_Ti} (a)-(c)). It should be noted that the temporary unavailability of the IVCP also contributed to elevating the overall divertor radiation, which is evident when comparing Figures \ref{fig:rad_Ti} (a)-(c) with Figures \ref{fig:ELMDetection} (d) and (e). In general, divertor gas puffing degrades global plasma performance \cite{paper:maggi2015pedestal}; the corresponding drop in temperature and rise in electron density increased localized radiation in the divertor region \cite{paper:pitcher1997experimental, paper:fournier2000calculation}. The evolution of the $T_{\mathrm{i}}$ profile during active $f_{\mathrm{ELM}}$ control (Figure \ref{fig:rad_Ti} (d)) is fully consistent with the temporal measurements in Figure \ref{fig:kinetics} (d). Crucially, these observations reveal an operational threshold. As established in Figure \ref{fig:ELMDetection}, the complete absence of divertor gas puffing leads to an exponential increase in $P_{\mathrm{core}}$. Conversely, applying excessive divertor gas yields a significant $\beta_{\mathrm{N}}$ drop without providing any substantial further reduction in $P_{\mathrm{core}}$ (Figures \ref{fig:control} and \ref{fig:kinetics}). This implies the existence of a critical minimum $f_{\mathrm{ELM}}$ threshold: exceeding this frequency merely degrades plasma performance without yielding additional impurity flushing benefits. Therefore, the developed real-time controller can be utilized to maintain $f_{\mathrm{ELM}}$ just above this critical threshold using minimal divertor gas puffing, thereby optimizing overall plasma performance while successfully preventing core W accumulation.

\section{Conclusion}
\label{sec:conclusion}
The ELM frequency ($f_{\mathrm{ELM}}$) was successfully controlled in real time using a proportional-integral (PI) controller with a divertor gas puff with $D_2$ gas as the actuator. This controller was designed based on the plasma response to divertor gas termination in shot 42160, an event that concurrently triggered an exponential increase in $P_{\mathrm{core}}$. In shot 43187, the controller adaptively modulated the divertor gas puff to accurately track a two-step $f_{\mathrm{ELM}}$ target. Comparisons against a reference discharge (shot 43186, featuring a 1.5 V feedforward gas puff) demonstrated that excessive divertor gas injection degrades global plasma performance without yielding any significant further reduction in $P_{\mathrm{core}}$. Synthesizing the results from these discharges confirms that active feedback control can maintain an optimal $f_{\mathrm{ELM}}$ threshold—one that is precisely sufficient to flush W from the core while avoiding the confinement degradation associated with excessive gas puffing. Moving forward, this controller can be utilized to optimize existing W-avoidance strategies and explore new operational regimes by identifying ideal $f_{\mathrm{ELM}}$ targets for more demanding tokamak scenarios.

In future studies, the experimental results presented here should be compared with discharges utilizing the active in-vessel cryopump (IVCP). Additionally, edge-localized resonant magnetic perturbations (ERMPs) \cite{paper:yang2024tailoring, paper:yang2025extending} can be employed as a complementary actuator to control $f_{\mathrm{ELM}}$. Ultimately, determining the optimal combination of divertor gas puffing and RMPs could maintain a sufficient $f_{\mathrm{ELM}}$ to flush out core W while simultaneously maximizing global plasma performance. Developing such real-time optimization schemes will become increasingly critical as KSTAR gradually upgrades its first wall and upper divertor to tungsten, complementing the currently installed lower tungsten divertor. Furthermore, to prepare for next-generation, high-power fusion devices, future research must explore integrated controllers that coordinate multiple actuators—such as divertor gas puffing, impurity seeding, and RMPs. Such a system must be capable of simultaneously (1) managing divertor heat flux via active detachment, (2) flushing core impurities, and (3) sustaining a grassy ELM regime.

\appendix

\section{Details of system identification and a PI controller design}
\label{appendix:control}

The tokamak serves as the plant for the control system. The system input, $u(t)$, is the voltage applied to the $\mathrm{D}_2$ divertor gas puff, and the system output, $y(t)$, is the ELM frequency, $f_{\mathrm{ELM}}(t)$. The plasma response is approximated using a first-order-plus-dead-time (FOPDT) model, which is governed by the following differential equation:

\begin{equation}
\frac{dy(t)}{dt} = -\frac{y(t) + K u(t-L)}{T}, 
\label{eq:dynamics}
\end{equation}

\noindent where $T$, $K$, and $L$ are real-valued parameters ($T, K, L \in \mathbb{R}$) representing the system time constant, steady-state gain, and dead time, respectively. With the model in Eq. (\ref{eq:dynamics}), the transfer function of the system, $H(s)$, is given by

\begin{align}
H(s) = \frac{Y(s)}{U(s)} = \frac{K}{sT+1}e^{-sL}.
\label{eq:transfer_laplace}
\end{align}

\noindent The impulse response of the system is given by an inverse Laplace transform of Eq. (\ref{eq:transfer_laplace}),

\begin{align}
h(t) = \frac{K}{T} e^{-(t-L)/T} \, (L \leq t).
\label{eq:impulse}
\end{align}

\begin{figure}[tbp]
\begin{center}
\includegraphics[width = 0.5 \linewidth]{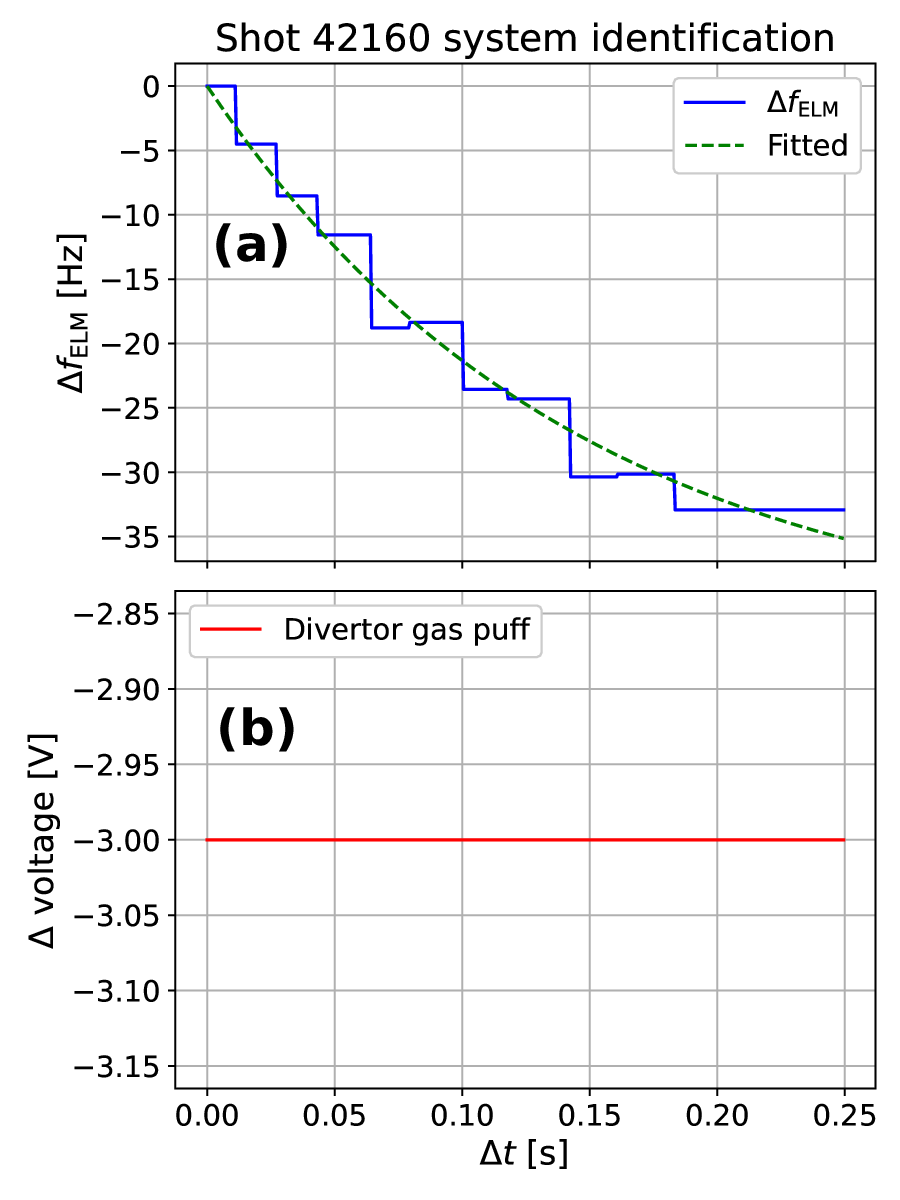}
\caption{
System identification results for KSTAR shot 42160. The system identification is performed over the interval from $t = 4$ to $5.35$ s. To extract the system gain ($K$) and time constant ($T$), the variations in the voltage command ($\Delta \mathrm{Voltage}$) and the ELM frequency ($\Delta f_{\mathrm{ELM}}$) between $t = 5.1$ and $5.35$ s are utilized, with the dead time ($L$) fixed at $1.1$ s. \textbf{(a)} The measured $\Delta f_{\mathrm{ELM}}$ from the plasma control system (PCS) compared against the fitted FOPDT model response. \textbf{(b)} The corresponding actuator step change ($\Delta \mathrm{Voltage}$) used to generate the fit.
}
\label{fig:SI}
\end{center}
\end{figure}

\noindent The system identification process is to find $K$, $T$, and $L$ in Eq. (\ref{eq:impulse}) which minimizes the error between the convolution integration

\begin{align}
&u(t) * h(t) \nonumber
\\ = &\int^t_0 u(\tau) \cdot h(t-\tau) d\tau, 
\label{eq:convolution}
\end{align}

\noindent and the output $y(t) = f_{\trm{ELM}}(t)$. KSTAR shot 42160 in Figure \ref{fig:ELMDetection} is chosen to find the coefficients in Eq. \ref{eq:transfer_laplace}. The divertor gas puff at the DH port was applied from 1 to 4 seconds with 3V, then turned off to observe the $f_{\trm{ELM}}$ response. The dead time, $L$, is set to 1.10, as the significant drop in $f_{\trm{ELM}}$ began after 1.1 seconds, as shown in Figure \ref{fig:ELMDetection} (b). The optimization process to find $K=14.3$ and $T=0.145$ was achieved by utilizing a Python package \texttt{scipy.optimize.curve\_fit}. The system identification result is shown in Figure \ref{fig:SI}.

A proportional-integral controller is adopted for our task, and its transfer function is given by

\begin{align}
C(s) = K_{\mathrm{p}}  + \frac{K_{\mathrm{i}}}{s}.
\label{eq:PID}
\end{align}

\noindent The proportional term looks at the current error for the fast response, while the integral term looks at past errors to achieve zero steady-state error. The derivative term looks at future errors to suppress overshooting, but it is not used because the system's response is slow and has a large dead time. With the Eq. (\ref{eq:transfer_laplace}) and Eq. (\ref{eq:PID}), the open-loop transfer function, $L(s)$, is calculated as 

\begin{align}
L(s) &= C(s) H(s) \nonumber 
\\ 
&= (K_{\mathrm{p}}  + \frac{K_{\mathrm{i}}}{s})(\frac{K}{sT+1}e^{-sL}).
\label{eq:loop}
\end{align}

\begin{figure}[tbp]
\begin{center}
\includegraphics[width = \linewidth]{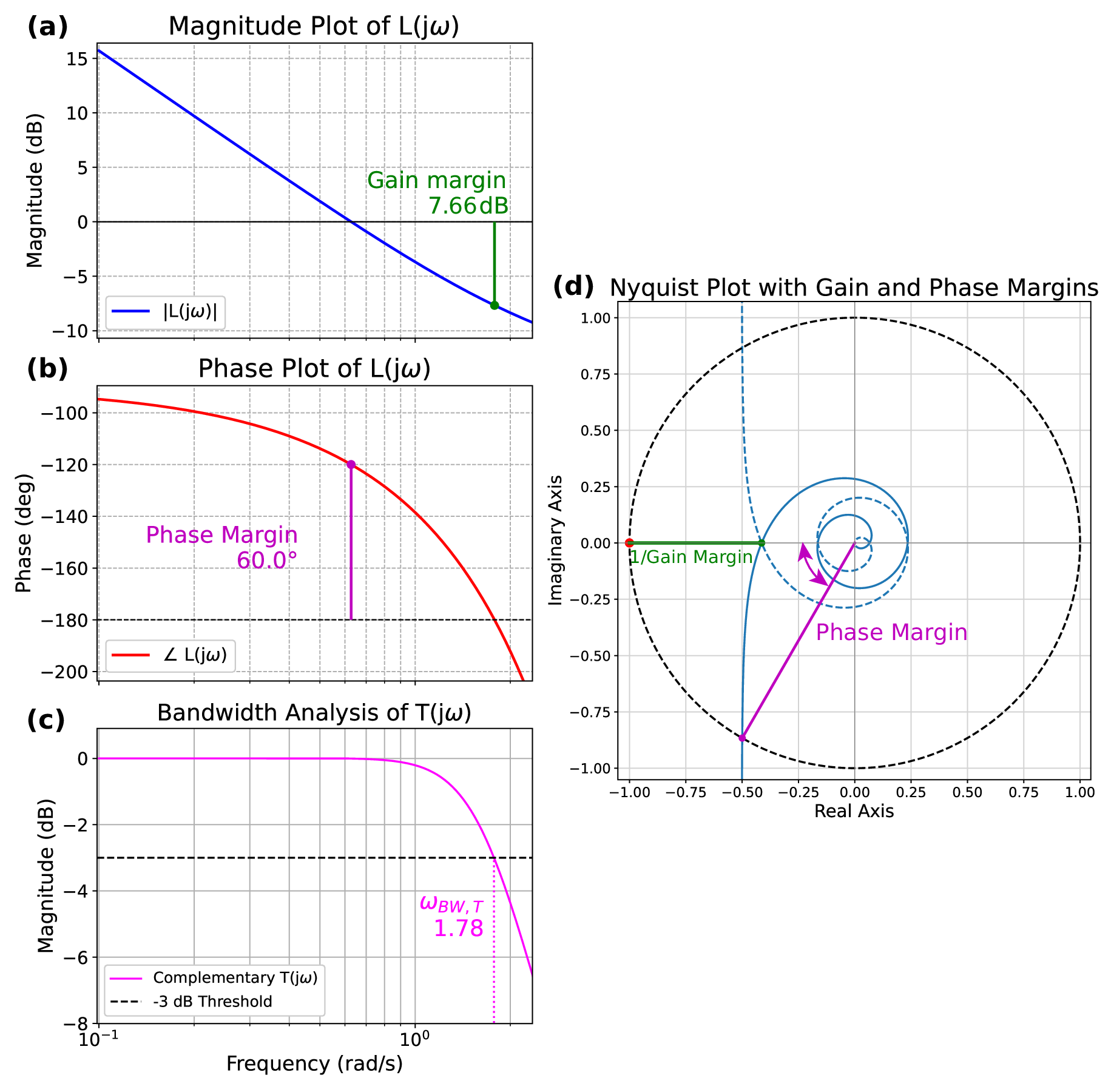}
\caption{
Frequency-domain stability and performance analysis of the PI controller designed for the divertor gas puff (PVB). Shown are the \textbf{(a)} Bode magnitude plot, \textbf{(b)} Bode phase plot, \textbf{(c)} magnitude plot of the complementary sensitivity function, and \textbf{(d)} Nyquist plot. The proportional and integral gains of the controller are $K_{\mathrm{p}} = 1.79 \times 10^{-2}$ and $K_{\mathrm{I}} = 4.26 \times 10^{-2}$, respectively. The resulting closed-loop system exhibits a phase margin of $60.0^\circ$ and a bandwidth of $\omega_{\mathrm{BW,T}} = 1.78$ rad/s.
}
\label{fig:Bode}
\end{center}
\end{figure}

The proportional and integral gains, $K_{\mathrm{p}}$ and $K_{\mathrm{i}}$, in Eq. (\ref{eq:loop}) were tuned to achieve a phase margin of approximately $60^\circ$ using the \texttt{pidtune} function in MATLAB. Figure \ref{fig:Bode} presents the Bode and Nyquist plots for the open-loop transfer function, $L(j\omega)$, alongside the complementary sensitivity function (equal to the closed-loop transfer function), $T(j\omega) = L(j\omega) / [1 + L(j\omega)]$. The optimized PI gains, the stability margins of $L(s)$, the bandwidth of $T(s)$, and the resulting closed-loop poles and zeros are summarized in Table \ref{tb:PID_bandwidth}.

\begin{table*}[tbp]
\centering
\caption{PI Controller summary. $p_1$ and $p_2$ are the closed-loop poles, and $z_1$ is a closed-loop zero.}

\renewcommand{\arraystretch}{1.2}
    \resizebox{\linewidth}{!}{%
    \begin{tabular}{ccccccccc}
    \toprule
    \midrule
    \textbf{Actuator} & \textbf{$K_p$} & \textbf{$K_I$} & $p_1$ & $p_2$ & $z_1$ & \textbf{Gain margin} & \textbf{Phase margin} &  \textbf{$\omega_{\trm{BW, T}}$}\\
    \midrule
    
    Divertor gas &  1.79 $\times$ $10^{-2}$ & 4.26 $\times$ $10^{-2}$ & -8.15 & -5.16 $\times$ $10^{-1}$ & -2.38 & 7.66 dB & 60.0$^{\circ}$ & 1.78 \\
     
    \midrule
    \bottomrule
    
    \label{tb:PID_bandwidth}
    \end{tabular}%
    }
\end{table*}

\ack{We sincerely appreciate the valuable discussions regarding experimental planning and data analysis with Qiming Hu, Seongmoo Yang, Cheolsik Byun, June-woo Juhn, Giwook Shin, Junhyeok Yoon, Azaraksh Jalalvand, and YongSu Na. We also thank Keith Erickson for establishing the initial setup for the PCS algorithm. The authors used Grammarly and Google's Gemini to support the language editing and revision of this manuscript.}

\funding{This work was supported by the U.S. Department of Energy, Office of Fusion Energy Sciences, under Awards DE-AC02-09CH11466 and DE-SC0024527. Support was also provided by the R\&D Program "High Performance Tokamak Plasma Research \& Development (EN2601-17)" and "Korea-US Collaboration Research for High Performance Plasma on Tungsten Divertor (EN2603-02)" through the Korea Institute of Fusion Energy (KFE), funded by the Government of the Republic of Korea.}


\data{
All data necessary to support the conclusions of this study are included within the article. However, the raw data is not publicly available due to institutional data policies of the Korea Institute of Fusion Energy (KFE).
}

\bibliographystyle{iopart-num}
\bibliography{ELM_freq_control}

\end{document}